%% file: main.tex
\documentclass[
]{ceurart}

\usepackage{listings}
\usepackage{subfig}

\usepackage{array}
\newcommand{\PreserveBackslash}[1]{\let\temp=\\#1\let\\=\temp}
\newcolumntype{C}[1]{>{\PreserveBackslash\centering}p{#1}}
\newcolumntype{R}[1]{>{\PreserveBackslash\raggedleft}p{#1}}
\newcolumntype{L}[1]{>{\PreserveBackslash\raggedright}p{#1}}
\newcolumntype{P}[1]{>{\raggedright\arraybackslash}m{#1}}

\begin{document}

\copyrightyear{2026}
\copyrightclause{Copyright for this paper by its authors.
Use permitted under Creative Commons License Attribution 4.0 International (CC BY 4.0).}

\conference{KDD 2026 AI for Education Day, at the ACM SIGKDD Conference on Knowledge Discovery and Data Mining (KDD~'26)}

\title{Engagement Intensity as a Learner-Modeling Signal for Adaptive AI Ethics Instruction}

\author[1]{Yongkyung Oh}[%
email=yongkyungoh@mednet.ucla.edu,
]
\author[1]{Lynn Talton}[%
email=ltalton@mednet.ucla.edu,
]
\author[1]{Alex Bui}[%
email=buia@mii.ucla.edu,
]
\cormark[1]
\address[1]{University of California, Los Angeles (UCLA), Los Angeles, CA, USA}

\cortext[1]{Corresponding author.}

\begin{abstract}
    Adaptive AI ethics instruction in graduate research training benefits from intake measures that reflect differences in prior LLM experience. Prior coursework or workshop attendance is an obvious candidate, but it is not clear whether it is associated with pre-instruction ratings on key AI perception items. We compare three candidate intake features, self-reported usage frequency, self-rated LLM familiarity, and prior AI education, across five baseline perception outcomes in 93 bioscience graduate and postdoctoral trainees enrolled in a required research ethics course. Usage frequency shows Holm-corrected associations with all five outcomes, self-rated familiarity with three, and prior AI education with none. A threshold-like pattern at the lower end of the scale is most visible for training interest and accuracy trust rather than appearing as a uniform gradient across all five outcomes. In a short intake survey, reported LLM use is more consistently associated with these perceptions than prior coursework or workshops, with self-rated familiarity serving as a secondary indicator. These results suggest that simple pre-instruction behavioral signals can inform lightweight intake profiling for adaptive AI ethics education.
\end{abstract}


\maketitle

\section{Introduction}\label{sec:intro}
Graduate trainees enter AI ethics instruction with different prior experiences and different views of what LLMs can do. LLMs introduce distinct challenges in higher education~\cite{milano_large_2023}, including graduate research. Generative AI tools are already used for research and writing support~\cite{roberts_artificial_2024,barrot_using_2023}, and many trainees use them before formal instruction.
This heterogeneity creates a design problem in higher-education settings already adapting to generative AI~\cite{bond_meta_2024,mcdonald_generative_2025}. A single AI ethics module may not serve learners with very different starting points equally well. AI ethics instruction for science and engineering trainees has already been implemented in graduate settings~\cite{usher_unpacking_2024}. The open question is which intake features best differentiate learners before that instruction begins.

Established AI literacy frameworks focus on learner-centered instruction for non-expert users~\cite{long_what_2020}, and recent meta-reviews highlight the need for more rigorous empirical work on how learners engage with AI in educational settings~\cite{bond_meta_2024}. Building on this perspective, we evaluate three practical intake measures that can be collected before instruction begins: self-reported usage frequency, self-rated LLM familiarity, and prior AI education. We compare their ability to differentiate learners at baseline and identify which measure provides the most informative signal for instructional personalization~\cite{xing_use_2025}.

Prior coursework and workshop attendance are intuitive proxies for readiness, but whether these proxies capture the variation that actually matters for instructional design has not been tested in this context. Recent guidance argues that responsible research use of AI depends on user judgment, oversight, and disclosure~\cite{resnik_ethics_2025,hosseini_ethics_2023}, but how to translate those principles into intake-level instructional design remains unclear. Existing AI ethics modules for graduate science and engineering trainees show measurable gains but treat learners as a relatively homogeneous group at intake~\cite{usher_unpacking_2024,borenstein_emerging_2021}.
Therefore, this study examines pre-instruction survey data from 93 bioscience graduate and postdoctoral trainees in a required Responsible Conduct of Research (RCR) course.

This paper makes two contributions. (1) We compare three candidate intake features for intake-level segmentation, namely usage frequency, self-rated LLM familiarity, and prior AI education. (2) We find that usage frequency is most consistently associated with baseline AI perceptions, self-rated familiarity is a weaker but meaningful secondary axis, and prior AI education does not survive family-wise correction, indicating that behavioral and self-perception signals are more informative intake features than education-history labels alone.


\section{Related Work}\label{sec:related_work}
Generative AI raises distinct pedagogical, methodological, and ethical challenges in higher education~\cite{kasneci_chatgpt_2023}. In doctoral populations, perceived usefulness and ease of use are also strong correlates of acceptance~\cite{zou_use_2023}.
A pre-post study of undergraduates reports shifts in cognitive, affective, and behavioral attitudes after sustained exposure to generative AI~\cite{suh_generative_2025}. Recent survey work also treats ChatGPT perceptions and usage as measurable constructs in higher-education and postgraduate contexts~\cite{ravselj_higher_2025,nemt-allah_validating_2024}. On the educator side, instructors in U.S.\ higher education report uneven and often cautious trust in generative AI, indicating that AI perceptions vary across roles and not only with individual experience~\cite{lyu_understanding_2025}. Calls to embed AI literacy and ethics into higher-education curricula have grown in parallel~\cite{xing_use_2025}.

\textbf{Prior Experience and AI Perception.}\;
Familiarity and perceived usefulness shape how people evaluate and adopt AI systems~\cite{kelly_what_2023,kasneci_chatgpt_2023}. 
A cross-national study of teacher trust in AI-based educational technology found that experiential variables such as AI self-efficacy and understanding carried more explanatory weight than demographic variables such as age and gender~\cite{viberg_what_2025}. In higher-education samples, similar patterns emerge. 
Trust in AI-powered educational technology correlates with demographic and academic characteristics among higher-education students~\cite{nazaretsky_critical_2025}.

\textbf{Trust Calibration in Human-AI Interaction.}\;
Trust calibration refers to the correspondence between a user's subjective trust and a system's objective capability~\cite{lee_trust_2004}.
Reviews of human trust in AI show that trust develops through system design and interaction context~\cite{glikson_human_2020}, while a recent review of human-AI interaction notes that the field still lacks a consistent definition of appropriate trust~\cite{mehrotra_systematic_2024}.
Users' knowledge, skills, and abilities also shape how they interpret AI outputs~\cite{alarcon_explicating_2025}.

\textbf{Adaptive Instruction and Learner Modeling.}\;
In intelligent tutoring, the choice of initial learner variable matters for adaptive instruction. VanLehn's review of intelligent-tutoring research shows that the match between learner starting point and instructional support shapes outcomes~\cite{vanlehn_relative_2011}. By analogy, we compare reported AI use, LLM familiarity, and prior AI education as alternative intake variables for the same cohort. Taken together, these studies point to experience-related differences, but they do not show whether reported AI use separates pre-instruction response patterns better than prior coursework or workshops in a graduate research setting.

\section{Study Design}\label{sec:study}

We use ``engagement intensity'' as a working label for the pair of intake features, reported use and self-rated familiarity, treated as complementary indicators of technology engagement in technology acceptance and AI literacy research~\cite{venkatesh_user_2003,kelly_what_2023}. Throughout the paper, ``adaptive'' refers to intake-level learner grouping based on pre-instruction signals rather than differentiated instructional materials, adaptive sequencing, personalized feedback, or a full intelligent tutoring system~\cite{vanlehn_relative_2011,xing_use_2025}.

This study was approved by the University of California, Los Angeles (UCLA) Institutional Review Board (IRB-25-1106). Participation was voluntary, consent was implied by beginning the anonymous survey, respondents could skip any item, and no identifying information was stored. We surveyed 93 bioscience graduate and postdoctoral trainees enrolled in a required Responsible Conduct of Research course and analyze only pre-instruction responses. Required RCR training is now standard in U.S.\ research settings, although its content and delivery vary across programs~\cite{steneck_history_2007,anderson_what_2007}. The descriptive intake distributions use the full cohort ($N=93$). Two respondents left all five focal outcome items blank, yielding an outcome-valid analytic sample of $N_{\text{valid}} = 91$ for all association tests. Data were collected with an anonymous online Qualtrics survey (UCLA Health, about 10 to 15 minutes) administered around the RCR course in a pre/post design, of which this pilot analyzes the pre-class responses only.

The sample is predominantly early-stage doctoral students drawn from more than ten bioscience subdisciplines. The baseline composition and disciplinary breakdown, the three intake-feature distributions, and reported LLM tools and purposes are detailed in Appendix~\ref{app:stats} (Tables~\ref{tab:composition} and~\ref{tab:engagement}).

\textbf{Measures}\;
Five perception items served as focal outcomes, each on a five-point Likert scale.
\begin{itemize}
\item Accuracy trust. ``I believe LLMs provide accurate general scientific information.''
\item Distinguishing capability. ``I feel capable of distinguishing between factual and incorrect information produced by LLMs.''
\item Complex-task trust. ``I trust LLMs for complex ethical issues or nuanced scientific concepts.''
\item Critical-thinking risk. ``I believe over-reliance on LLMs might impair my critical thinking skills.''
\item Training interest. ``I am interested in receiving formal training to effectively utilize LLMs in my future research projects.''
\end{itemize}

Each item is a single five-point Likert statement (anchored from ``Strongly disagree'' to ``Strongly agree'') and is analyzed as a distinct facet of pre-instruction AI perception rather than pooled into a composite. No internal-consistency or factor-structure statistic is reported, because the items are treated as distinct facets. Each item corresponds to a construct examined in prior work, namely trust in factual accuracy~\cite{lee_trust_2004,glikson_human_2020}, self-rated ability to evaluate AI output~\cite{long_what_2020}, trust calibration for complex tasks~\cite{mehrotra_systematic_2024,alarcon_explicating_2025}, perceived over-reliance risk~\cite{zhai_effects_2024}, and interest in formal training~\cite{borenstein_emerging_2021,mccoy_training_2025}. 

\textbf{Analysis}\;
Because Likert responses are ordinal rather than interval~\cite{jamieson_likert_2004}, usage frequency and self-rated LLM familiarity were tested as five-level ordinal predictors using Spearman correlations~\cite{spearman_proof_1904}. Prior AI education, a single categorical item covering training in AI, computational biology, or bioinformatics, was tested with Kruskal--Wallis $H$ tests~\cite{kruskal_use_1952} under the three-group coding. 

We additionally report a collapsed two-group coding (any prior versus none, with group sizes 51 and 40). Because the Kruskal--Wallis test reduces to the Mann--Whitney $U$ test for two groups, the same test serves both codings, and Cliff's $\delta$ summarizes the two-group effect size. Holm correction~\cite{holm_simple_1979} was applied across the five outcomes within each predictor family, because each family was treated as a competing intake signal rather than pooled into a single 15-test omnibus family.

\section{Results}\label{sec:results}

\begin{figure}[!htb]
\centering
\includegraphics[width=0.90\linewidth]{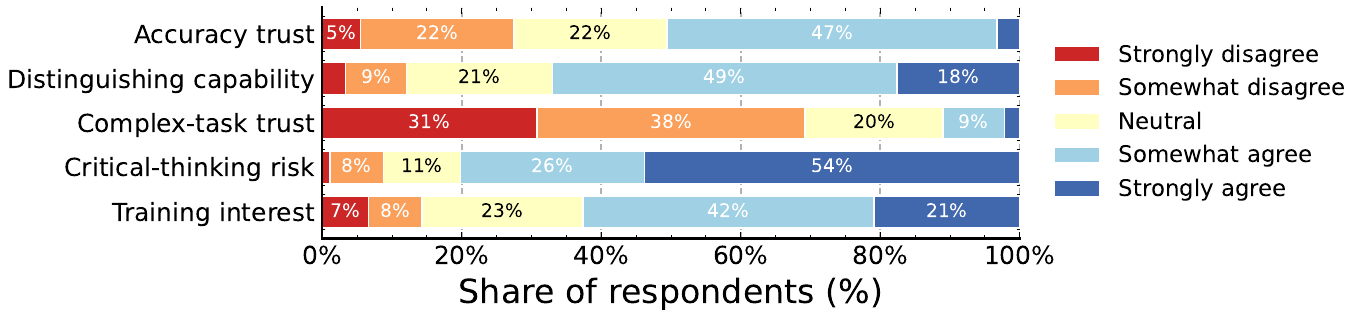}
\caption{Response-profile distribution across the five focal perception outcomes. The stacked profile shows how agreement, neutrality, and disagreement are distributed within each outcome.}
\label{fig:response_profile}
\end{figure}

Figure~\ref{fig:response_profile} summarizes the overall response distributions for the five focal outcomes. The profile shows high endorsement of critical-thinking risk and training interest, moderate confidence in distinguishing capability, and substantially more cautious trust for complex-task use than for general scientific accuracy.

Tables~\ref{tab:usage},~\ref{tab:familiarity}, and~\ref{tab:education} summarize the three candidate intake features side by side. Usage frequency survives Holm correction on all five outcomes, self-rated familiarity on three, and prior AI education on none. Usage frequency shows the largest associations for complex-task trust ($\rho=.409$), accuracy trust ($\rho=.370$), and critical-thinking risk ($\rho=-.346$). Self-rated familiarity shows a weaker pattern and survives Holm correction for distinguishing capability ($\rho=.294$, $p_{\mathrm{Holm}}=.023$), complex-task trust ($\rho=.278$, $p_{\mathrm{Holm}}=.031$), and critical-thinking risk ($\rho=-.254$, $p_{\mathrm{Holm}}=.045$). Accuracy trust shows a positive familiarity association but does not survive Holm correction ($p_{\mathrm{Holm}}=.069$), and training interest is not associated with familiarity. The familiarity pattern is bottom-anchored rather than smoothly graded. Separation is clearest at the lower end of the scale, while the medians for the ``Somewhat familiar'' and ``Very familiar'' groups are often identical. In this sample, the five pre-instruction items vary more by reported use and familiarity than by prior AI education. 

\input{tables/out_usage}
\input{tables/out_familiarity}
\input{tables/out_education}

Table~\ref{tab:education} isolates prior AI education, the one intake feature with no family-wise association. Under the three-group coding (formal / informal / none), the Kruskal--Wallis tests are non-significant for all five outcomes after Holm correction (largest $H=5.06$), and the sample is dominated by the None category (44.1\%). The null is robust to coding granularity. A collapsed two-group contrast (any prior versus none), reported in the same table, reaches the identical conclusion (minimum $p_{\mathrm{Holm}}=.524$) with at most small effect sizes (Cliff's $|\delta|<.19$). The largest two-group gap is a one-level median difference on accuracy trust and on critical-thinking risk, both non-significant. Prior coursework or workshop labels therefore separate baseline AI perceptions less sharply than reported use behavior or self-rated familiarity, an informative null that sharpens, rather than weakens, the intake-feature hierarchy.

\section{Discussion}\label{sec:discussion}
Usage frequency showed Holm-corrected associations on all five outcomes, self-rated familiarity on three, and prior AI education on none. Under this coarse three-category coding, workshop or course participation alone did not distinguish pre-instruction ratings as clearly as the two experience-related features.
In Appendix~\ref{app:stats}, Figure~\ref{fig:experience_ladders} shows this pattern most clearly. The early threshold-like pattern is strongest for training interest and accuracy trust, but it is not uniform across all five outcomes. A practical distinction in introductory design may therefore lie between learners who have not used AI tools and those with any sustained use, while still recognizing that some outcomes, such as critical-thinking risk, shift mainly among daily users.

The negative association between usage and critical-thinking risk ($\rho = -.346$) adds a calibration concern. Systematic review evidence suggests that over-reliance on AI dialogue systems can affect students' cognitive performance~\cite{zhai_effects_2024}. In the present sample, heavy users reported less worry about over-reliance, which could reflect appropriate confidence or reduced sensitivity to failure modes that grow harder to detect in more capable models~\cite{zhou_larger_2024}. The familiarity result points in the same direction. Learners who felt more familiar with LLMs also reported lower concern about critical-thinking risk ($\rho = -.254$). Taken together, these results raise a calibration concern among more engaged learners, who may need exercises that focus on when LLM outputs fail.

Why behavioral engagement tracks these perceptions is not established here, but two mechanisms are plausible. Repeated interaction may recalibrate accuracy trust as users encounter system failures~\cite{lee_trust_2004,glikson_human_2020}, and for the over-reliance item the negative association could instead reflect growing confidence or habituation rather than heightened concern~\cite{zhai_effects_2024,zhou_larger_2024}. Hands-on use may also increase self-rated evaluation competence rather than measured skill. We treat these as candidate accounts to be tested longitudinally rather than as established pathways.

For intake profiling, the current data support using reported usage frequency as a first intake variable and self-rated familiarity as a secondary check. If a course uses simple routing, the present results are more consistent with a split between non-users and sustained users than with a split based on prior coursework or workshops. Heavy users may also need additional calibration-focused activities. Furthermore, we discuss alternative non-causal accounts of these associations in Appendix~\ref{app:alt}.

\section{Conclusion}\label{sec:conclusion}
Understanding what learners bring into AI ethics instruction has become a pressing concern as generative AI use spreads across graduate research. Among the three intake features examined, usage frequency shows the strongest association with the five pre-instruction perception items, self-rated familiarity shows a weaker pattern, and prior AI education does not reach significance after correction. A practical implication is that short pre-instruction surveys can use simple behavioral and self-perception signals to inform intake profiling, with reported LLM use indicating engagement level and self-rated familiarity serving as a secondary check on perceived readiness. In heterogeneous graduate cohorts, intake analytics could, in principle, help distinguish learners who need basic orientation from those who benefit from calibration-focused work on trust, over-reliance, and failure modes.

\textbf{Limitations.}\; This study is cross-sectional, associational, and drawn from a single institution. The single-respondent ``Not at all familiar'' group means the familiarity ladder should be interpreted descriptively and not overclaimed. 
Both predictors and outcomes are single-item self-report measures, so common-method variance cannot be excluded and multi-item reliability is unavailable. Other engagement dimensions remain outside the present analysis. Whether differentiated instruction based on these profiles improves learning outcomes requires an intervention study.

\textbf{Future Work.}\; Future work should test whether this hierarchy of intake features replicates across institutions, disciplines, and course formats. The engagement construct could also be broadened with behavioral measures such as session logs or task-level interaction data. Randomized studies could then evaluate whether engagement-based grouping improves calibration, learning, or transfer.

\begin{acknowledgments}
  This study was approved by the University of California, Los Angeles (UCLA) Institutional Review Board (IRB-25-1106). 
  We thank the trainees who participated in the Spring 2025 Responsible Conduct of Research (RCR) course survey.

\end{acknowledgments}

\section*{Declaration on Generative AI}
During the preparation of this work, the authors used generative AI tools for grammar and spelling checking and writing-style improvement. After using these tools, the authors reviewed and edited the content as needed and take full responsibility for the content of this publication.

{\small
\bibliography{references}
}

\clearpage
\newpage
\appendix

\section{Sample Composition and Supporting Visualizations}\label{app:stats}

Table~\ref{tab:composition} reports the baseline composition. The sample is predominantly early-stage doctoral students (77.4\%), with a modal age of 25 to 29 (52.7\%) and a slight female majority (53.8\%). Participants span more than ten bioscience subdisciplines, with neuroscience and molecular biology the largest at 16.1\% each and no field exceeding 17.0\%, so no single subfield drives the results. 

\input{tables/composition}

Table~\ref{tab:engagement} reports the generative-AI engagement profile. The three intake features differ in shape, which bears on their resolution as predictors. Reported usage spreads fairly evenly across the five levels, self-rated familiarity concentrates at the upper end (about 79.6\% Somewhat or Very familiar, with a single Not-at-all respondent), and prior AI education is dominated by the None category (44.1\%). Reported tool use is dominated by ChatGPT (86.0\%), and the most common purposes are searching for scientific information (52.7\%) and summarizing literature (40.9\%).

\input{tables/engagement}

\begin{figure}[!htb]
\centering
\subfloat[Usage-frequency gradient\label{fig:ladder_usage}]{\includegraphics[width=0.85\linewidth]{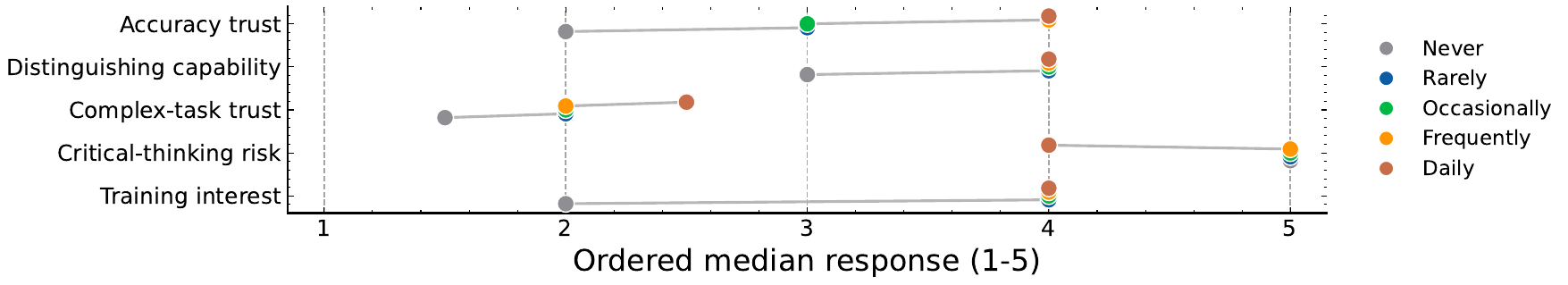}}
\\
\subfloat[LLM-familiarity gradient\label{fig:ladder_familiarity}]{\includegraphics[width=0.85\linewidth]{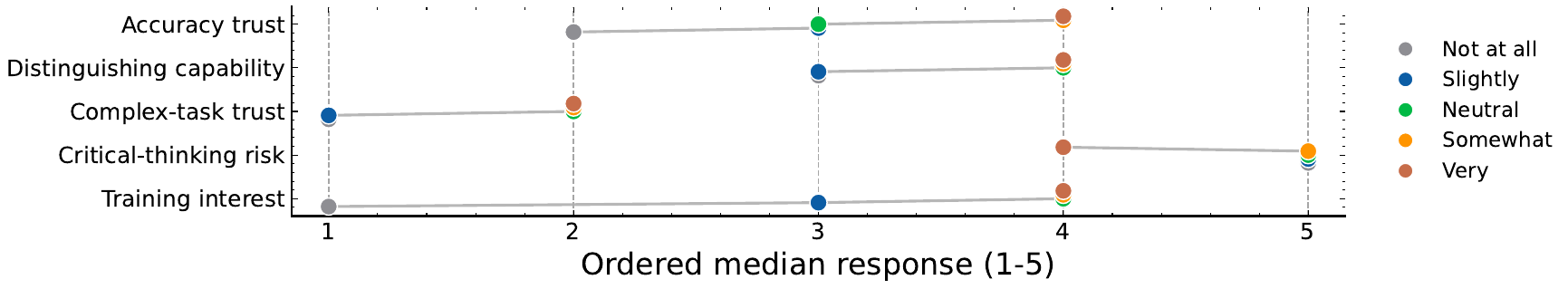}}
\\
\subfloat[Prior-education boundary (three-group)\label{fig:ladder_education}]{\includegraphics[width=0.85\linewidth]{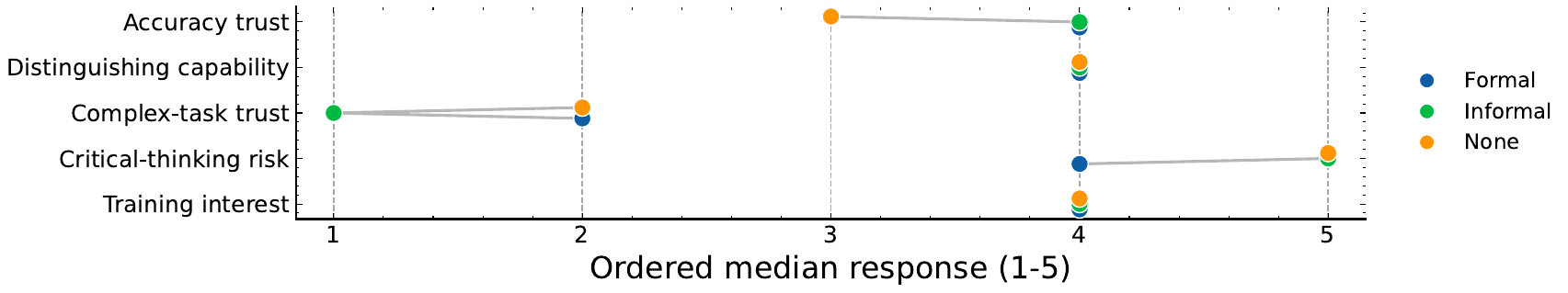}}
\\
\subfloat[Prior-education boundary (two-group)\label{fig:ladder_education_2group}]{\includegraphics[width=0.85\linewidth]{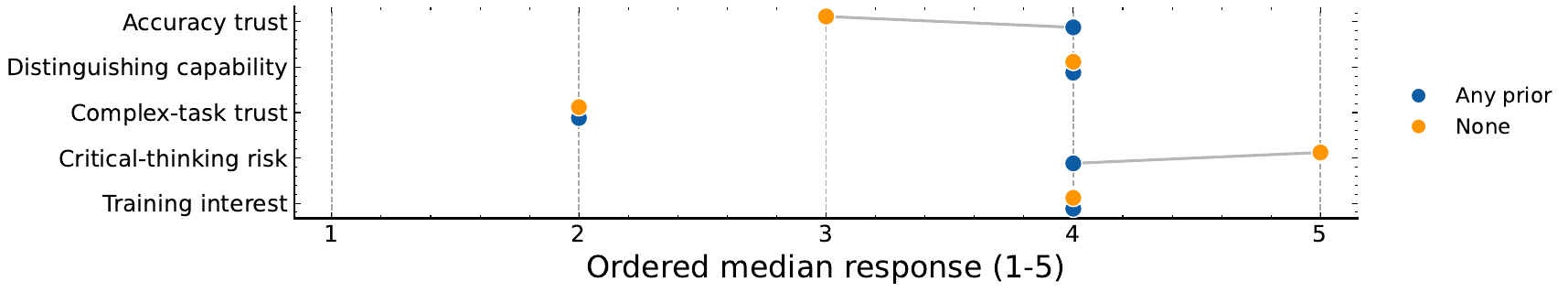}}
\caption{Ordered-median ladders for accuracy trust, distinguishing capability, complex-task trust, critical-thinking risk, and training interest across the candidate intake features. Panels (c) and (d) show prior AI education under the three-group (formal / informal / none) and two-group (any prior / none) codings.
}
\label{fig:experience_ladders}
\end{figure}

Figure~\ref{fig:experience_ladders} visualizes the gradients described in Section~\ref{sec:results}. Usage frequency shows the strongest pattern, particularly for training interest and accuracy trust. Familiarity shows weaker gradients for evaluation capability, complex-task trust, and critical-thinking risk, although the ``Not at all familiar'' group contains only one respondent. Prior AI education shows little separation under either coding scheme, consistent with its null result. Overall, the patterns distinguish non-users from learners with sustained engagement but do not support a uniform binary threshold across outcomes.

\section{Alternative Interpretations}\label{app:alt}

The study design does not establish directionality. Engagement may influence perceptions, perceptions may influence engagement, or both may reflect upstream factors such as disciplinary norms or dispositional openness~\cite{venkatesh_user_2003,glikson_human_2020}. The usage measure also captures self-reported frequency on a Never-to-Daily scale rather than observed behavior, making it an indirect measure of engagement~\cite{nemt-allah_validating_2024}.

Part of the observed associations may reflect general response styles, such as acquiescence or extremity bias, rather than differences in knowledge or readiness, because both predictors and outcomes were measured through self-report~\cite{podsakoff_common_2003}.
The cross-sectional design cannot distinguish between these possibilities. The five outcomes were treated as facets of distinct constructs documented in prior work on trust calibration and AI literacy, namely accuracy trust, evaluation capability, complex-task trust, over-reliance risk, and training interest, rather than as indicators of a single latent dimension~\cite{lee_trust_2004}. 

The null result for prior AI education should be interpreted cautiously. AI literacy spans multiple competencies~\cite{long_what_2020}, so the three-category coding (Formal, Informal, None) cannot capture the depth, recency, or duration of training and is coarser than the five-level usage and familiarity scales. However, a higher-powered two-group comparison (any prior education versus none) produced the same null result, with at most small effects. This pattern suggests that the limitation lies in the measurement of prior AI education rather than in the construct itself. Future measures should capture the depth, recency, and duration of training.

\end{document}

%% file: tables/out_usage.tex
\begin{table}[htb]
\caption{Usage-frequency associations with the five focal outcomes. For each outcome the table reports the median rating within every usage level from Never to Daily, together with the Spearman correlation $\rho$ and its Holm-corrected $p$ value computed across the five outcomes. 
}\label{tab:usage}
\centering\scriptsize
\begin{tabular}{l C{1.4cm}C{1.4cm}C{1.4cm}C{1.4cm}C{1.4cm} cc}
\toprule
& \multicolumn{5}{c}{Median by usage level} & \multicolumn{2}{c}{Test statistics} \\
\cmidrule(lr){2-6}\cmidrule(lr){7-8}
Outcome & Never & Rarely & Occasionally & Frequently & Daily & $\rho$ & $p_{\mathrm{Holm}}$ \\
\midrule
Accuracy trust         & 2.0 & 3.0 & 3.0 & 4.0 & 4.0 &  .370   & .001**      \\
Distinguishing capability & 3.0 & 4.0 & 4.0 & 4.0 & 4.0 &  .281   & .014*       \\
Complex-task trust     & 1.5 & 2.0 & 2.0 & 2.0 & 2.5 &  .409   & {<}.001***  \\
Critical-thinking risk & 5.0 & 5.0 & 5.0 & 5.0 & 4.0 & $-.346$ & .002**      \\
Training interest      & 2.0 & 4.0 & 4.0 & 4.0 & 4.0 &  .215   & .041*       \\
\midrule
\textit{n} ($N_{\text{valid}}=91$) & 10 & 13 & 21 & 27 & 20 & & \\
\bottomrule
\end{tabular}
\end{table}

%% file: tables/out_familiarity.tex
\begin{table}[htb]
\caption{Self-rated LLM-familiarity associations with the five focal outcomes. For each outcome the table reports the median rating within every familiarity level from Not at all to Very, together with the Spearman correlation $\rho$ and its Holm-corrected $p$ value computed across the five outcomes. 
}\label{tab:familiarity}
\centering\scriptsize
\begin{tabular}{l C{1.4cm}C{1.4cm}C{1.4cm}C{1.4cm}C{1.4cm} cc}
\toprule
& \multicolumn{5}{c}{Median by familiarity level} & \multicolumn{2}{c}{Test statistics} \\
\cmidrule(lr){2-6}\cmidrule(lr){7-8}
Outcome & Not at all & Slightly & Neutral & Somewhat & Very & $\rho$ & $p_{\mathrm{Holm}}$ \\
\midrule
Accuracy trust         & 2.0 & 3.0 & 3.0 & 4.0 & 4.0 &  .222   & .069   \\
Distinguishing capability & 3.0 & 3.0 & 4.0 & 4.0 & 4.0 &  .294   & .023*  \\
Complex-task trust     & 1.0 & 1.0 & 2.0 & 2.0 & 2.0 &  .278   & .031*  \\
Critical-thinking risk & 5.0 & 5.0 & 5.0 & 5.0 & 4.0 & $-.254$ & .045*  \\
Training interest      & 1.0 & 3.0 & 4.0 & 4.0 & 4.0 &  .042   & .694   \\
\midrule
\textit{n} ($N_{\text{valid}}=91$) & 1 & 13 & 5 & 47 & 25 & & \\
\bottomrule
\end{tabular}
\end{table}

%% file: tables/out_education.tex
\begin{table}[htb]
\caption{Prior-education associations with the five focal outcomes under a three-group coding of formal, informal, and none and a collapsed two-group coding of any prior versus none. The table reports the group medians together with the Kruskal--Wallis $H$ statistic, its Holm-corrected $p$ value, and Cliff's $\delta$ as the two-group effect size. 
}\label{tab:education}
\centering\scriptsize
\begin{tabular}{l C{1.1cm}C{1.1cm}C{1.1cm} C{1.1cm}C{1.1cm} ccccc}
\toprule
& \multicolumn{3}{c}{Median (three-group)} & \multicolumn{2}{c}{Median (two-group)} & \multicolumn{5}{c}{Test statistics} \\
\cmidrule(lr){2-4}\cmidrule(lr){5-6}\cmidrule(lr){7-11}
Outcome & Formal & Informal & None & Any prior & None & $H_3$ & $p_3$ & $H_2$ & $p_2$ & $\delta$ \\
\midrule
Accuracy trust         & 4.0 & 4.0 & 3.0 & 4.0 & 3.0 & 2.38 & .576 & 2.32 & .524 &    .17 \\
Distinguishing capability & 4.0 & 4.0 & 4.0 & 4.0 & 4.0 & 3.80 & .576 & 2.63 & .524 &    .18 \\
Complex-task trust     & 2.0 & 1.0 & 2.0 & 2.0 & 2.0 & 3.88 & .576 & 1.12 & .724 &$-$.12 \\
Critical-thinking risk & 4.0 & 5.0 & 5.0 & 4.0 & 5.0 & 5.06 & .397 & 1.37 & .724 &$-$.13 \\
Training interest      & 4.0 & 4.0 & 4.0 & 4.0 & 4.0 & 3.56 & .576 & 0.49 & .724 &    .08 \\
\midrule
\textit{n} ($N_{\text{valid}}=91$) & 32 & 19 & 40 & 51 & 40 & & & & & \\
\bottomrule
\end{tabular}
\end{table}

%% file: tables/composition.tex
\begin{table}[htb]
\caption{Baseline sample composition for the full cohort of 93 trainees ($N=93$).}
\label{tab:composition}
\centering\scriptsize
\begin{minipage}[t]{0.49\linewidth}
\centering
\begin{tabular}[t]{L{5.7cm}R{1.1cm}}
\toprule
\textbf{Demographics} & \textit{n} (\%) \\
\midrule
\multicolumn{2}{l}{\textit{Academic stage}} \\
\quad PhD student, year 1--2              & 72 (77.4) \\
\quad PhD student, year 3 or above           & 7 (7.5) \\
\quad Postdoctoral researcher                & 6 (6.5) \\
\quad Dual-degree student (MD/PhD)            & 6 (6.5) \\
\quad Other trainee          & 2 (2.2) \\
\multicolumn{2}{l}{\textit{Gender}} \\
\quad Female                 & 50 (53.8) \\
\quad Male                   & 39 (41.9) \\
\quad Prefer not to say      & 4 (4.3) \\
\multicolumn{2}{l}{\textit{Age}} \\
\quad 18--24                 & 32 (34.4) \\
\quad 25--29                 & 49 (52.7) \\
\quad 30--34                 & 7 (7.5) \\
\quad 35--39                 & 3 (3.2) \\
\quad 40 or older            & 2 (2.2) \\
\bottomrule
\end{tabular}
\end{minipage}
\hfill
\begin{minipage}[t]{0.49\linewidth}
\centering
\begin{tabular}[t]{L{5.7cm}R{1.1cm}}
\toprule
\textbf{Discipline} & \textit{n} (\%) \\
\midrule
Molecular Biology & 15 (16.1) \\
Neuroscience & 15 (16.1) \\
Genetics \& Genomics & 11 (11.8) \\
Cell \& Developmental Biology & 9 (9.7) \\
Physics \& Biology in Medicine & 8 (8.6) \\
Biochemistry & 7 (7.5) \\
Microbiology \& Immunology & 7 (7.5) \\
Pharmacology & 7 (7.5) \\
Bioinformatics & 4 (4.3) \\
Physiology & 4 (4.3) \\
Biomathematics & 1 (1.1) \\
Chemistry & 1 (1.1) \\
Environmental \& Molecular Toxicology & 1 (1.1) \\
Medical Informatics & 1 (1.1) \\
Psychology & 1 (1.1) \\
Structural Biology & 1 (1.1) \\
\bottomrule
\end{tabular}
\end{minipage}
\end{table}

%% file: tables/engagement.tex
\begin{table}[htb]
\caption{Generative-AI engagement profile ($N=93$). Tools used and purposes permit multiple responses (percentages of $N=93$). Usage, familiarity, and education are single-response.}\label{tab:engagement}
\centering\scriptsize
\begin{minipage}[t]{0.49\linewidth}
\centering
\begin{tabular}[t]{L{5.7cm}R{1.1cm}}
\toprule
\textbf{LLM usage} & \textit{n} (\%) \\
\midrule
\multicolumn{2}{l}{\textit{Tools used}} \\
\quad ChatGPT                & 80 (86.0) \\
\quad Google Gemini          & 21 (22.6) \\
\quad Microsoft Copilot      & 12 (12.9) \\
\quad Perplexity             & 11 (11.8) \\
\quad Claude                 & 3 (3.2) \\
\quad Other                  & 2 (2.2) \\
\multicolumn{2}{l}{\textit{Purposes for use}} \\
\quad Search for scientific information & 49 (52.7) \\
\quad Summarize literature/readings   & 38 (40.9) \\
\quad Understand complex concepts    & 36 (38.7) \\
\quad Assist with writing papers & 24 (25.8) \\
\quad Identify knowledge gaps & 23 (24.7) \\
\quad Translate technical texts        & 19 (20.4) \\
\quad Generate ideas/questions         & 17 (18.3) \\
\quad Create outlines/drafts        & 11 (11.8) \\
\bottomrule
\end{tabular}
\end{minipage}
\hfill
\begin{minipage}[t]{0.49\linewidth}
\centering
\begin{tabular}[t]{L{5.7cm}R{1.1cm}}
\toprule
\textbf{AI background} & \textit{n} (\%) \\
\midrule
\multicolumn{2}{l}{\textit{Usage frequency}} \\
\quad Never                  & 10 (10.8) \\
\quad Rarely                 & 13 (14.0) \\
\quad Occasionally           & 22 (23.7) \\
\quad Frequently             & 27 (29.0) \\
\quad Daily                  & 21 (22.6) \\
\multicolumn{2}{l}{\textit{LLM familiarity}} \\
\quad Not at all familiar    & 1 (1.1) \\
\quad Slightly familiar      & 13 (14.0) \\
\quad Neutral / Not sure     & 5 (5.4) \\
\quad Somewhat familiar      & 48 (51.6) \\
\quad Very familiar          & 26 (28.0) \\
\multicolumn{2}{l}{\textit{Prior AI education}} \\
\quad Formal                 & 33 (35.5) \\
\quad Informal               & 19 (20.4) \\
\quad None                   & 41 (44.1) \\
\bottomrule
\end{tabular}
\end{minipage}
\end{table}